\documentclass[12pt,preprint]{aastex}
\usepackage{epsfig}
\usepackage{lscape}

\makeatletter

\newcommand{\Rmnum}[1]{\expandafter\@slowromancap\romannumeral #1@}
\makeatother

\def\Lsun{\hbox{\it L$_\odot$}}

\def\Msun{\hbox{\it M$_\odot$}}

\newcommand{\etal}{\mbox{et al.}}

\newcommand{\program}[1]{{\tt {#1}}}
\def\program{\texttt}

\begin{document}
\title{Discovery of a Luminous Blue Variable with an Ejection Nebula Near the Quintuplet Cluster}

\shorttitle{}

\author{J.~ C. Mauerhan\altaffilmark{1}, M.~R. Morris\altaffilmark{2},  A. Cotera\altaffilmark{3}, H. Dong\altaffilmark{4}, Q.~D. Wang\altaffilmark{4}, S.~R. Stolovy\altaffilmark{1}, C. Lang\altaffilmark{5}, I. S. Glass\altaffilmark{6} }
\altaffiltext{1}{Spitzer Science Center, California Institute of Technology, Mail Code 220-6, 1200 East California Blvd., Pasadena, CA 91125, USA; mauerhan@ipac.caltech.edu} 
\altaffiltext{2}{Department of Physics and Astronomy, University of California, Los Angeles, CA 90095, USA}
\altaffiltext{3}{SETI Institute, 515 North Whisman Road, Mountain View, CA 94043, USA}
\altaffiltext{4}{Department of Astronomy, University of Massachusetts, Amherst, MA 01003, USA} 
\altaffiltext{5}{Department of Physics and Astronomy, University of Iowa, Iowa City, IA 52245, USA}
\altaffiltext{6}{South African Astronomical Observatory, P.O. Box 9, Observatory 7935, South Africa}

\begin{abstract}
We report on the discovery of a luminous blue variable (LBV) lying $\approx$7 pc in projection from the Quintuplet cluster. This source, which we call LBV G0.120$-$0.048, was selected for spectroscopy owing to its detection as a strong source of Paschen-$\alpha$ (P$\alpha$) excess in a recent narrowband imaging survey of the Galactic center region with the \textit{Hubble Space Telescope}/Near-Infrared Camera and Multi-Object Spectrometer. The $K$-band spectrum is similar to that of the Pistol Star and other known LBVs. The new LBV was previously cataloged as a photometric variable star, exhibiting brightness fluctuations of up to $\approx$1 mag between 1994 and 1997, with significant variability also occurring on month-to-month time scales. The luminosity of LBV G0.120$-$0.048, as derived from Two Micron All Sky Survey photometry, is approximately equivalent to that of the Pistol Star. However, the time-averaged brightness of LBV G0.120$-$0.048 between 1994 and 1997 exceeded that of the Pistol Star; LBV G0.120$-$0.048 also suffers more extinction, which suggests that it was intrinsically more luminous in the infrared than the Pistol Star between 1994 and 1997. P$\alpha$ images reveal a thin circular nebula centered on LBV G0.120$-$0.048 with a physical radius of $\approx$0.8 pc. We suggest that this nebula is a shell of ejected material launched from a discrete eruption that occurred between 5000 and 10,000 years ago. Because of the very short amount of time that evolved massive stars spend in the LBV phase, and the close proximity of LBV G0.120$-$0.048  to the Quintuplet cluster, we suggest that this object might be coeval with the cluster, and may have once resided within it. 

 \end{abstract}
\section{Introduction}
Luminous blue variables (LBVs) are descendants of massive O stars which are nearing the end of core hydrogen burning.  The extremely high luminosities of these stars ($L_{\textrm{\scriptsize{bol}}}\sim10^5$--$10^7$ \Lsun) place them near or above the Eddington limit, resulting in large variability amplitudes and violent mass ejections. As heavy mass-loss ensues, the outer layers of hydrogen are lost, exposing deeper layers of the star, at which point these objects evolve into Wolf-Rayet stars (Langer et al. 1994). LBVs are extremely rare; only $\approx$10 have been confirmed in the Milky Way. Some known Galactic LBVs include P Cygni, $\eta$ Car, AG Car, LBV 1806$-$20, the Pistol star, and  qF362.  The latter two LBVs reside within the Quintuplet starburst cluster near the Galactic center (Figer et al. 1995; Geballe et al. 2000), and their presence among $\sim$100 O stars and dozens of more highly evolved, coeval Wolf--Rayet stars in the cluster provides insight as to how LBVs fit into the overall picture of massive star evolution. In this Letter, we report on the discovery of an LBV lying near the Quintuplet cluster, which we will refer to as LBV G0.120$-$0.048, based on its Galactic coordinates.  

\section{Observations}
\subsection{Selection of LBV G0.120$-$0.048}
LBV G0.120$-$0.048 was highlighted following a Paschen-$\alpha$ (P$\alpha$) narrowband imaging survey of the Galactic center region (Wang et al. 2010), performed with the \textit{Hubble Space Telescope} (\textit{HST}) and the Near-Infrared Camera and Multi-Object Spectrometer (NICMOS). The survey covered a total projected area of 2253 pc$^2$ in 144 orbits of observations. The NIC3 camera was utilized and narrowband images were obtained through the F187N (line) and F190N (continuum) filters. Dozens of unidentified point sources of P$\alpha$ line excess were discovered in the survey area, including LBV G0.120$-$0.048. Recent spectroscopic identifications of other P$\alpha$ emission-line stars will be presented in a forthcoming paper. LBV G0.120$-$0.048  was immediately recognized as an intriguing object, owing to the large P$\alpha$ excess measured for this star, which is comparable to, but exceeds that of the Pistol Star, and its close proximity to the Quintuplet starburst cluster.  The position and photometry of the central point source is presented in Table 1, along with data for the other Quintuplet LBVs qF362 and the Pistol Star for comparison. The $JHK_s$ photometry included in Table 1 is from the Two-Micron All Sky Survey (2MASS) catalog (Cutri et al. 2003), and the date of the 2MASS measurements was Julian day (JD) 2451825.4954.   	 

Another remarkable feature of LBV G0.120$-$0.048 is the circular nebula of P$\alpha$ emission surrounding it, shown in Figure 1. The nebula has an angular radius of $\approx$200{\arcsec}, and has no counterpart in the adjacent F190N continuum, which indicates that the emission is entirely from the P$\alpha$ transition. 

\subsection{Spectroscopy}
A $K$-band spectrum of LBV G0.120$-$0.048  was obtained on 2008 May 15, using the 4.1 m Anglo-Australian Telescope and the IRIS2 instrument (Tinney et al. 2004). The sky was clear and the average seeing was 1\farcs5. The slit on IRIS2 is 1\arcsec~wide, providing a resolution of $R\approx2400$ in the $K$ band. The telescope was nodded in an ABBA sequence for sky subtraction and bad-pixel suppression. The data images were reduced and the spectrum was extracted  using the \program{Starlink ORACDR} pipeline. A telluric spectrum was obtained from observations of the A0V star HD155379, and was applied to the science data using the IDL-based program \program{xtellcor\_general} (Vacca et al. 2003).

\section{Analysis}
\subsection{Near-infrared Spectra}
The $K$-band spectrum of LBV G0.120$-$0.048  is presented in Figure 2, and has a signal-to-noise ratio of $\approx$50--80, moving from the red end to the blue end of the spectrum. The spectrum is dominated by Br$\gamma$ emission at $\lambda$2.167 {\micron}, which appears as a P Cygni profile. He {\sc i} emission is detected at $\lambda$2.06 {\micron} and $\lambda$2.150 {\micron}. Two weak absorption features of He {\sc i} might also be detected blueward of  the Br$\gamma$ line.  Low-ionization emission features are also present, including Mg {\sc ii} at $\lambda$2.137 and $\lambda$2.144 {\micron}, Na {\sc i} at $\lambda$2.206 and $\lambda$2.209 {\micron}, and relatively strong Fe {\sc ii} emission at $\lambda$2.089 {\micron}. Weak forbidden emission features of iron ([Fe {\sc ii}]) are also detected at $\lambda$2.133 {\micron}, and perhaps in a blend with He {\sc i} at $\lambda$2.043 {\micron}. Detailed information on the atomic transitions detected in the spectrum and their equivalent widths are presented in Table 2; line centers and equivalent widths were measured using the \program{splot} routine in \program{IRAF}.   Based on the presence and relative strengths of these emission features, the spectrum of LBV G0.120$-$0.048 is similar to those of the LBV stars WRA 751, AG Car, the Pistol Star, and the B[e] star GG Car (Morris et al. 1996; Figer et al. 1998).

We used the Br$\gamma$ P Cygni profile to obtain an estimate for the wind velocity of LBV G0.120$-$0.048. The absorption trough is centered at $\lambda$2.1674 {\micron} and the emission peak is centered at  $\lambda$2.1684 {\micron}, which implies a wind velocity of $\approx$140 km s$^{-1}$. This value is comparable to the wind velocities derived for the Pistol Star and qF362, which have terminal values of $\approx$105 km s$^{-1}$ and $\approx$170 km s$^{-1}$, respectively (Najarro et al. 2009). However, the measured wind velocity of LBV G0.120$-$0.048 is a lower limit on the terminal velocity, since it is only relevant for the region in the wind where the H {\sc i} absorption occurs. It is possible that additional radiative acceleration of the wind continues at radii further from the star than this region.

\subsection{Extinction and Luminosity}
To derive the absolute photometry of LBV G0.120$-$0.048,  we must first correct for interstellar extinction. In order to compare the photometry of LBV G0.120$-$0.048  with that of the Pistol star and qF362 in a consistent way, we adopted the same intrinsic color, bolometric correction, and extinction law used by Figer et al. (1998) for the analysis of the Pistol Star, using $(H-K)_0=-0.05$, $A_K=1.78E_{H-K}$, and $BC_K=-1.5$ mag. We assumed that the extinction relation holds for the $K_s$ band as well, and used the 2MASS photometry from Table 1 to compute the extinction values for LBV G0.120$-$0.048, the Pistol Star, and qF362. The derived $K_s$-band extinction values for LBV G0.120$-$0.048, the Pistol Star, and qF362 are $A_{K_s}= 3.26, 2.99$, and 3.42 mag, respectively. Assuming a distance of 8 kpc to the Galactic center (Reid 1993), this implies absolute $K_{s}$ magnitudes of
$M_{K_s}=-10.3$, $-10.2$, and $-10.8$ mag for LBV G0.120$-$0.048, the Pistol Star, and qF362, respectively. Thus, the bolometric luminosities of LBV G0.120$-$0.048  and the Pistol Star are $L_{\textrm{\scriptsize{bol}}}=10^{6.6}$~{\Lsun} and $L_{\textrm{\scriptsize{bol}}}=10^{6.5}$~{\Lsun}, respectively. The derived luminosity for qF362 is higher, with$L_{\textrm{\scriptsize{bol}}}=10^{6.9}$~{\Lsun}. However, Najarro et al. (2009) derived stellar luminosities for the Pistol Star and qF362 that are significantly lower than the values presented here, deriving $L_{\textrm{\scriptsize{bol}}}\approx10^{6.2}$~{\Lsun} for both stars. Their analysis is based on the detailed modeling of the near-infrared spectra of these stars, which probably resulted in more accurate results than our photometry based calculations. Admittedly, our assumption of a universal bolometric correction and intrinsic color for LBVs is crude, and may have led to a significant error in our derived bolometric luminosities. Currently, the only inference we can make with confidence is that the relative $M_{K_s}$ values reflect relative $K_s$ luminosities.  Photometric monitoring of flux and color changes, in tandem with spectroscopic monitoring and modeling of the spectra, is necessary to elucidate the relationship between intrinsic infrared brightness, color, and bolometric luminosity.

\section{Discussion} 
\subsection{Variability of LBV G0.120$-$0.048} LBV G0.120$-$0.048  is a variable star, identified as such by Glass et al. (2001, 2002), who monitored the central $24{\arcmin}\times24{\arcmin}$ of the Galaxy in a $K$-band photometric campaign between 1994 and 1997, with about four individual observations per year. LBV G0.120$-$0.048, the Pistol Star, and qF362 were all detected as large-amplitude variables in this survey, cataloged with the respective designations 10-1, 13-4, and 13-6. The light curves of these three stars are presented in Figure 3; the data for the Pistol Star and qF362 were first presented in Glass et al. (1999).  LBV G0.120$-$0.048 appears to have undergone an overall decrease in brightness by $\approx$1 mag between 1994 and 1997, but also showed signs of significant intra-month variability ($\approx$0.35 mag) during 1994. During the same 4 year time span, the Pistol Star's variations were also significant ($\pm0.5$ mag or so) but  less extreme than LBV G0.120$-$0.048, while the magnitude of qF362's variations are similar to those of LBV G0.120$-$0.048. The time-averaged brightness and the variability amplitude of LBV G0.120$-$0.048 exceed that of the Pistol Star and qF362, which have $K$-band magnitudes and standard deviations of 6.86 (0.41), 7.38 (0.15), and 7.43 (0.26) mag, respectively. The fact that LBV G0.120$-$0.048 has a larger average brightness than the Pistol Star, while suffering more extinction (see Section 3.2), implies that it was more intrinsically more luminous in the infrared than the Pistol star throughout the duration of the Glass et al. (2001, 2002) survey. However, during the later 2MASS observation, which occurred on JD2451825.4954 (2000 October 17), the infrared luminosity of LBV G0.120$-$0.048 and the Pistol Star were more-or-less equivalent, while that of LBV qF362 exceeded both of them, as indicated by the analysis in Section 3.2.  Without color or spectroscopic information to accompany the flux changes in Figure 3 it is not possible to determine whether the variations reflect a change in the total bolometric luminosity of these stars, as appears to be the case for the LBV AG Car, or changes in spectral morphology at constant bolometric luminosity (e.g., see Groh et al. 2009 and references therein). Alternatively, infrared photometric variability may occur as a result of the variable free-free component induced by a changing mass-loss rate.  Again, we are in need of simultaneous photometric and spectroscopic monitoring to discriminate between potential causes for the brightness variations observed from these stars.

\subsection{The Nebular Shell of LBV G0.120$-$0.048}
The shell of P$\alpha$ emission surrounding LBV G0.120$-$0.048 is remarkably circular (presumably spherical), relative to the elongated morphology of the Pistol Nebula. We found no obvious counterpart to the shell of  LBV G0.120$-$0.048 in the \textit{Spitzer}/Infrared Array Camera mid-infrared images of the region, which contrasts with the Pistol Nebula, which is bright at mid-infrared wavelengths. Thus, it is unlikely that the shell surrounding LBV G0.120$-$0.048 is illuminated interstellar material, unrelated to the central LBV. Rather, the nebula appears to have been ejected from LBV G0.120$-$0.048. This would not be surprising, since the Pistol Nebula, and the nebulae surrounding other LBVs, such as  $\eta$ Carina, exhibit unambiguous kinematical evidence for ejection (e.g., see Figer et al. 1999b; Currie et al. 1996).  The $\approx$200{\arcsec} angular radius of the shell surrounding LBV G0.120$-$0.048 implies a physical radius of $\approx$0.8 pc. The average expansion velocity of the 10 LBV ejection nebulae, summarized in Table 6 of Figer et al. (1999b), is $\approx$100 km s$^{-1}$, with a standard deviation of $\approx$170 km s$^{-1}$; and most of these nebulae have estimated dynamical times of $\lesssim10^4$ yr. So, assuming this is close approximation of the expansion velocity for the shell surrounding LBV G0.120$-$0.048 implies that the presumed ejection event probably occurred $\approx$5000--10,000 yr before present.  

The new P$\alpha$ image in Figure 1 also reveals a noteworthy feature associated with the Pistol Nebula. The flattening of the Nebula's left side was suggested by Figer et al. (1999b) to be the result of either (1) the interaction of the Pistol Nebula with the stellar winds of Quintuplet cluster members, or (2) the interaction of the Pistol Nebula with the strong milligauss magnetic field that comprises the arched radio filaments, which are concentrated near the Quintuplet cluster. It has been suggested that the strong magnetic field that creates the filaments is pervasive throughout the Galactic center region (Yusef-Zadeh \& Morris 1987). The magnetic pressure from such a strong field will suppress the flow of ionized material, such as an expanding LBV ejection nebula,  in the direction perpendicular to the field, which may responsible for the flattening of the left side of the Pistol Nebula in Figure 1.  However, the faint rim of the Pistol Nebula's \textit{right} side is not flattened, which suggests that if the Galactic center magnetic field is responsible for the flattening of the left side of the Pistol Nebula, then the strong magnetic field may be a \textit{local} phenomenon, rather than a pervasive feature of the Galactic center region as a whole. 

Finally, we note that there may be evidence for a past outburst from LBV qF362 as well. Low-level P$\alpha$ emission, in the form of irregular knots and tendrils, appears to surround the position of qF362, apparent in Figure 1. This could be material thrown off by the star during a prior outburst. Alternatively, since qF362 appears to lie relatively close to the edge of the Sickle, the low-level emission surrounding it could be the ionized interstellar material, formerly of the Sickle structure, still in the process of being evaporated away. Deeper imaging of the low-level emission at higher resolution with the \textit{HST}/NICMOS NIC1 camera could resolve this issue by providing unambiguous evidence for a physical association between LBV qF362 and the low-level nebular emission surrounding it.

\subsection{Origin of LBV G0.120$-$0.048}
LBV G0.120$-$0.048 lies $\approx$2\farcm8 from the Quintuplet cluster, which also contains qF362 and the Pistol Star. Assuming a distance of 8 kpc to the Galactic center (Reid 1993), this angular separation implies a projected physical separation of $\approx$7 pc from the Quintuplet. The presence of three LBVs within such a small projected volume makes this region the largest known concentration of LBVs. Such a concentration is remarkable, given the very short period of time massive stars spend in the LBV phase ($\sim10^5$ yr, Langer et al. 1994). Therefore, we suggest that LBV G0.120$-0.048$ is coeval with the Quintuplet cluster, and either formed outside the main cluster, but during the same burst of star formation, or formed within the cluster and was subsequently ejected from it somehow.

Figer et al. (1998) suggested that the Pistol Star has an initial mass in the range of 150--250 \Msun. Owing to their similarities in intrinsic brightness and spectral properties, it would be reasonable to assume that LBV G0.120$-$0.048 and the Pistol Star have comparable masses. If LBV G0.120$-$0.048 did originate within the Quintuplet, it is hard to explain how such a massive star could be so far removed from the cluster, rather than residing near its center as a consequence of dynamical mass segregation. However, it has been suggested that LBVs might result from the merger of a close binary (Pasquali et al. 2000). Close binaries are expected to form in the dense environs of starburst clusters, while the same multi-body interactions that harden close binaries and induce mergers, also impart a systemic velocity to the hardened binary or merger byproduct that is on the order of 10 km s$^{-1}$ (Gaburov et al. 2010). Since the internal velocity dispersion of the Quintuplet is also $\sim$10 km s$^{-1}$ (Figer et al. 1999a), an additional velocity of 10 km s$^{-1}$ imparted to a star or binary could allow it to escape the cluster. The stellar merger origin has been proposed for the Pistol Star (Figer \& Kim 2002; Gaburov et al. 2008), while the dynamical ejection of hardened binaries produced within the Quintuplet was proposed to explain several colliding-wind binaries found near the Quintuplet (Mauerhan et al.  2007; Mauerhan et al. 2010).  Although the merger-byproduct hypothesis is currently a very speculative suggestion, it could be given credence if one or all of the Quintuplet LBVs were shown to be moving away from the cluster. Alternatively, LBV G0.120$-$0.048 may have formed during the same burst of star formation that created the Quintuplet, originating from the same molecular cloud, but never becoming a bound member of the cluster.  A proper motion measurement, enabled by adaptive optics, could provide evidence for or against a Quintuplet cluster origin within a few years.

\setlength{\tabcolsep}{0.06in}
\renewcommand{\arraystretch}{1.05}
\begin{landscape}
\begin{deluxetable}{llrrccccc}
\tablecolumns{9}
\tablewidth{0pc}
\tabletypesize{\scriptsize}
\tablecaption{Basic Data for LBV G0.120$-$0.048, the Pistol Star, and qF362}
\tablehead{
\colhead{Star} & \colhead{2MASS}& \colhead{$\textrm{R.A.}$} & \colhead{$\textrm{Decl.}$} &  \colhead{$J$} & \colhead{$H$} & \colhead{$K_s$} & \colhead{F190N} & \colhead{F187N/F190N} \\ [2pt]
\colhead{} & \colhead{Designation} & \multicolumn{2}{c}{(deg, J2000)}& \colhead{(mag)} & \colhead{(mag)} & \colhead{(mag)} & \colhead{(\textrm{\scriptsize{mJy}})} & \colhead{}}
\startdata
LBV G0.120$-$0.048 & J17460562$-$2851319 &   266.523436    &    $-$28.858866  & $12.53\pm0.03$ &$9.24\pm0.02$&$7.46\pm0.02$  & 1056$\pm$17 & $1.41\pm0.03 $\\                   
The Pistol Star             & J17461524$-$2850035 & 266.563502    &    $-$28.834328   & $11.83\pm0.02$ &$8.92\pm0.04$&$7.29\pm0.02$ &  1138$\pm$21 &$ 1.29\pm0.04$ \\               
 qF362                        & J17461798$-$2849034 &  266.574927     &    $-$28.817627   & $12.31\pm0.03$ & $8.97\pm0.03$ & $7.09\pm0.03$ &     853$\pm$20  & $1.10\pm0.04$
\enddata
\tablecomments{Positions and $JHK_s$ photometry were taken from the 2MASS catalog (Cutri et al. 2003). The date of $JHK_s$ measurements was JD2451825.4954. The narrowband F187N and F190N measurements are from H. Dong et al. (2010, in preparation).}
\end{deluxetable}
\end{landscape}

\begin{deluxetable}{lcccc}
\tablecaption{Spectral Lines Detected in LBV G0.120$-$0.048 \label{tbl-2}}
\tablewidth{0pt}
\tablehead{
\colhead{Ion} & \colhead{Transition} & \colhead{$\lambda$} &
\colhead{$\lambda_{\textrm{\scriptsize{obs}}}$} & \colhead{$W_{\lambda}$} \\
\colhead{} & \colhead{ ($l$--$u$)} & \colhead{($\mu$m)} & \colhead{($\mu$m)} & 
\colhead{(${\buildrel _{\circ} \over {\mathrm{A}}}$)}  
}
\footnotesize
\startdata
He {\sc i}      &  $6p^3P$--$4s^3S$                                           &  2.043        & 2.042    &  1.0 \\
~~+$[$Fe {\sc ii}$]$?     &  $a^4P_{5/2}$--$a^2P_{3/2}  $               &  2.047        & \nodata  &  \nodata \\
He {\sc i}      &  $2s^1S$--$2p^1P^0$                                           &  2.058        & 2.057    &  1.7 \\
Fe {\sc ii}       & $z^{4}F_{3/2}$--$c^{4}F_{3/2}$             &  2.089        & 2.087    &  1.5 \\
He {\sc i}            & $3p^{3}P$--$4s^{3}S + 3p^{1}P$--$4s^{1}S$     &  2.112        & 2.112    & $-$0.4 \\
$[$Fe {\sc ii}$]$          & $a^{2}P_{3/2}$--$a^{4}P_{3/2}$      &  2.134        & 2.134   & 0.3 \\
Mg {\sc ii}           & $5s^{2}S_{1/2}$--$5p^{2}P_{3/2}$           &  2.137        & 2.138    &  1.2 \\
Mg {\sc ii}           & $5s^{2}S_{1/2}$--$5p^{2}P_{1/2}$           &  2.144        & 2.144    &  1.0 \\
He {\sc i}           &     $4p^{3}P$--$7s^{3}S$                                      &  2.150        & 2.149     &  0.3 \\
H {\sc i}             & 4 -- 7                                        &  2.167        & 2.168    &  6.2 \\
Na {\sc i}            & $4p^{2}P^0_{3/2}$--$4s^{2}S_{1/2}$       &  2.206    & 2.208    &  3.7 \\
Na {\sc i}            & $4p^{2}P^0_{1/2}$--$4s^{2}S_{1/2}$       &  2.209     & 2.211    &  2.2 \\
\enddata
\tablecomments{~Line centers were adopted from Morris et al. (1996). Errors in equivalent width (EW) are $\approx$0.1--0.2 ${\buildrel _{\circ} \over {\mathrm{A}}}$. Negative values of EW are assigned to absorption lines; all other lines are in emission.}
\end{deluxetable}

\newpage

\begin{figure*}[t]
\centering
\epsscale{1}
\plotone{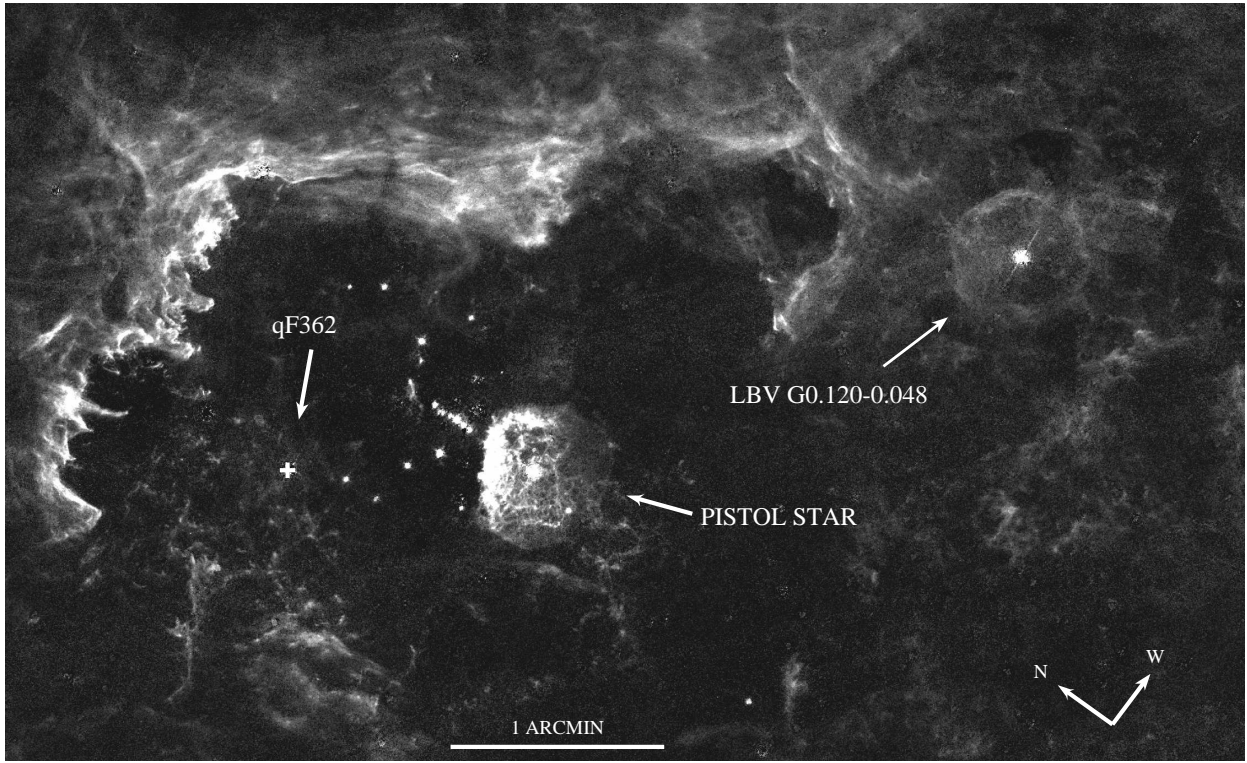}
\caption[Paschen alpha]{\linespread{1}\normalsize{\textit{HST/NICMOS} P$\alpha$ (F187N$-$F190N) survey image of the Quintuplet Cluster region (Wang et al. 2009). LBV G0.120$-$0.048 is visible in the upper right of the image as a bright point source surrounded by a spherical shell of nebulosity, which was presumably ejected from the star. The Pistol Star, surrounded by its bright ejection Nebula, lies to the right of the main cluster. The LBV qF362 (\textit{cross point}) also appears to be surrounded by some low-level nebular emission.}}
\label{fig:hardint}
\end{figure*}

\begin{figure*}[t]
\centering
\epsscale{1}
\plotone{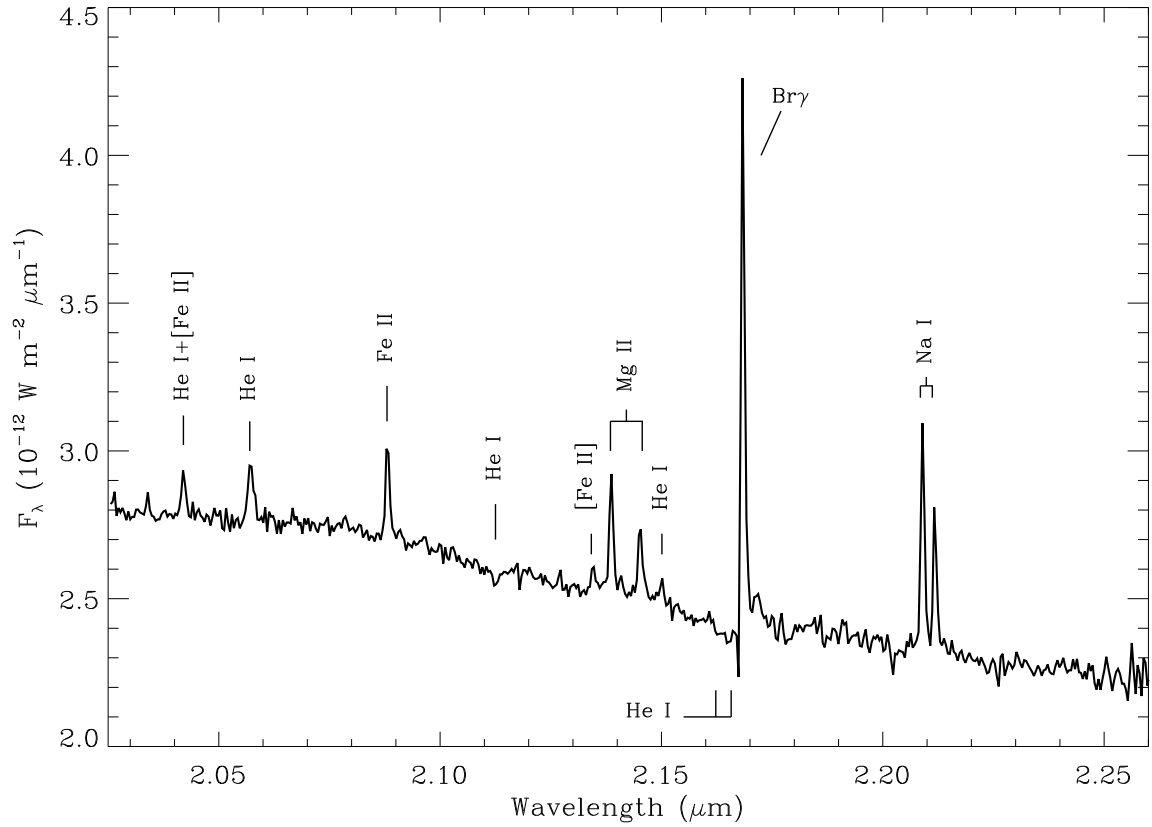}
\caption[LBV Spectrum]{\linespread{1}\normalsize{$K$-band spectrum of LBV G0.120$-$0.048, exhibiting emission features characteristic of known LBV and B[e] stars (see the text).}}
\end{figure*}

\begin{figure*}[t]
\centering
\epsscale{1}
\plotone{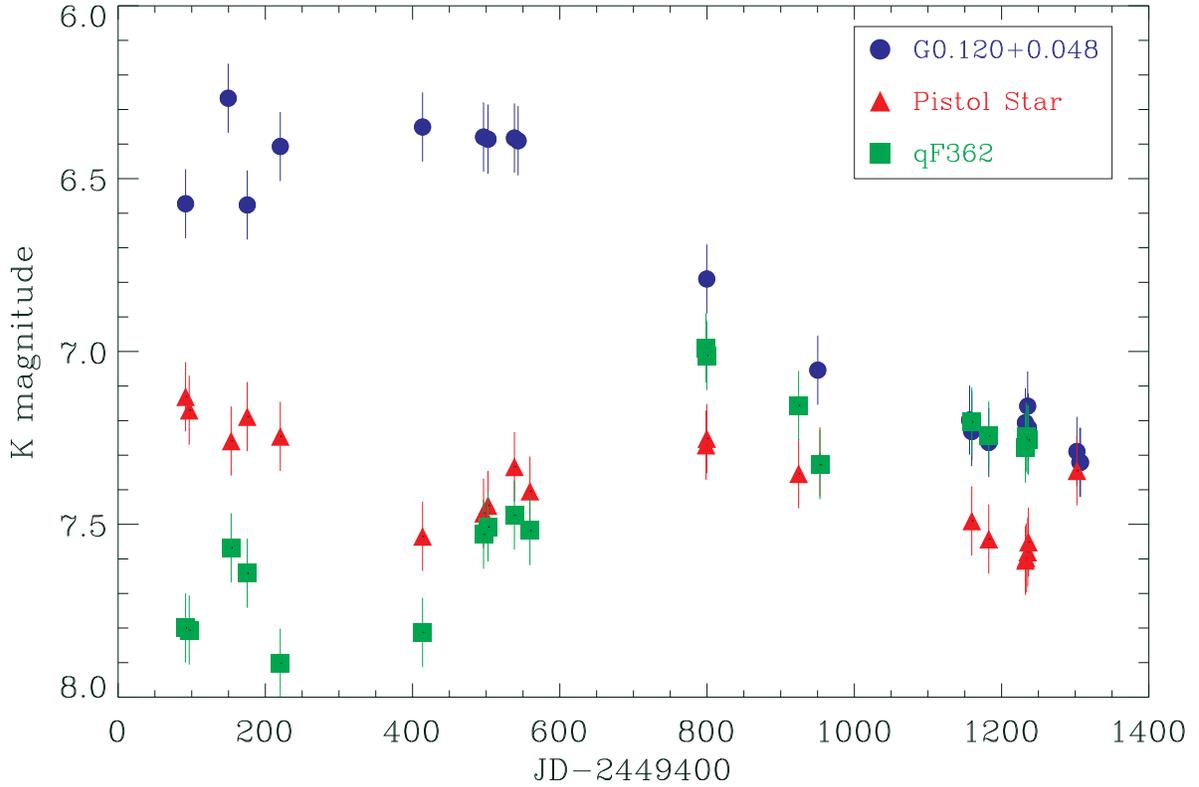}
\caption[]{\linespread{1}\normalsize{$K$-band light curves for LBV G0.120$-$0.048, the Pistol Star, and qF362. Photometric uncertainties are 0.1 mag, as derived using the nightly standard deviations in magnitude for standard stars used in Glass et al. (1999, 2001, 2002). The data for LBV G0.120$-$0.048 extends between 1994 May 19 and 1997 September 15, but are presented in Julian days for the figure.}}
\label{archesquintuplet}
\end{figure*}

\end{document}